\documentclass[twocolumn,showpacs,preprintnumbers,amsmath,amssymb]{revtex4}
\usepackage[dvips]{graphics}
\begin{document}
\title{Charged Plate Beyond Mean-field:  One-loop Corrections by Salt-density Fluctuation}
\author{William Kung}\email{wkung@physics.syr.edu}
\affiliation{Department of Physics, Syracuse University, Syracuse, New York 13244}
\author{A.W.C. Lau}
\affiliation{Department of Physics, Florida Atlantic University, Boca Raton, Florida 33431}
\date{\today}

\begin{abstract}
We present an exact field-theoretic formulation for a fluctuating, generally asymmetric, salt density in the presence of a charged plate.  The non-linear Poisson-Boltzmann equation is obtained as the saddle-point of our field theory action.  Focusing on the case of symmetric salt, we systematically compute first-order corrections arising from electrolytes fluctuation to the free energy density, which can be explicitly obtained in closed form.  We find that for systems with low to moderate salt density, fluctuation corrections to the free-energy depends sensitively on the salt concentration as well as their charge valency.  Further, we find that electrolyte fluctuation leads to a reduced electrostatic repulsion between two point-charges when they are close to the charged plate.  
\end{abstract}

\maketitle

\def\coth{\hbox{\rm{coth}}}
\def\csch{\hbox{\rm{csch}}}

\section{introduction}

Across systems from astronomical scale to cellular level, electrostatic interaction plays an important role in their characterization by thermal properties.  For these systems, electrostatic interactions are modulated by charged ions interacting via Coulomb and hard-core potentials.  Examples include high-density plasma found in stellar interior and solar photosphere~\cite{Netz00a, Alfven}, charged colloids found in such industrial applications as coating materials and ceramic precursors~\cite{Kung02, RusselText}, and the self-assembly of charged surfactants into lamellar phases~\cite{Parsegian06, Parsegian93}.  Water itself can be considered a weak electrolytic solution at room temperature, and as a result, considerations of electrostatic interactions figure prominently into many biological problems involving macroions in aqueous solutions~\cite{Israel}.  As examples, the macroions may be charged cell membranes or stiff polyelectrolytes such as DNA and polymers~\cite{PhysicsToday, NetzReview}.  Often, the interplay between thermal fluctuation and electrostatic interaction in these charged systems determine their chemical affinity, structural and functional properties.  To ensure their stability, there must also be overall charge neutrality.  In experimental conditions, the charged particles in aqueous solutions often exist in the form of dissociated cations and anions.    Due to the long-range nature of Coulomb interactions,  these charged particles can become correlated and drive the vastly complex behaviors observed in these systems.   

Quantitatively, it is assumed that the cationic and anionic densities in these charged systems are conserved and obey the Boltzmann distribution in equilibrium~\cite{RusselText}.  The conventional description then relies upon the mean-field approach in which one averages out the fluctuation in the distribution of salt ions.  Combining Gauss' law and the Boltzmann distribution of the salt ions then leads to the well-known Poisson-Boltzmann equation~\cite{VerweyText}.  The Poisson-Boltzmann equation provides a good quantitative description of most experimental systems and simulations~\cite{NetzReview, Travesset05} since in general each charged ion simultaneously interacts with many other ions in solution;  one can reproduce the interaction as an effective mean-field experienced by each ion throughout the system.  Despite the straightforward derivation of the Poisson-Boltzmann equation, exact solutions are only known in a few cases, when one imposes further simplifications such as assuming only the presence of counter-ions, restricting to problems possessing planar~\cite{Guoy1910, Chapman1913,  Stern1924} or cylindrical geometries~\cite{VerweyText, Fuoss1951}, and linearization of the Poisson-Boltzmann equation in the appropriate regime~\cite{Schiessel2003}.  

However, it is well known by now that the mean-field approach of the Poisson-Bolzmann equation works only in the low-charge regime~\cite{NetzReview,Schiessel2003,Shklovskii00, Shklovskii01, Liu99}.  There has been great interest to go beyond the mean-field description of the electrostatic systems, since the mean-field nature of the Poisson-Boltzmann equation necessarily precludes correlation effects.  The need to consider correlation arises, for example, in such problems as DNA condensation with multivalent cations~\cite{Shklovskii00, Shklovskii01, Liu99} where the macroions are highly charged and the correlation effects between ions become important.  Phenomena such as charge inversion and like-charge attraction~\cite{Yu2002, Levin2002,MoreiraText, Ha1997,Pincus1998, Lukatsky1999, Lau2002} have been reported in these cases.  Much experimental effort has been spent to confirm these predictions.  Experimental techniques involving manipulating optical tweezers in confined geometries~\cite{Squires2000,Behrens2001,Behrens2}, interaction measurements in equilibrium~\cite{Kepler1994,Vondermassen1994,Tinoco1996,Bongers1998,Brunner2002}, and digital video microscopy~\cite{Grier2004} have been used to measure attraction between like-charges in various systems.  

Aside from the importance of correlation effects which the mean-field approach, by construction, fails to incorporate, it has also been shown theoretically that within the framework of the Poisson-Boltzmann theory, the interaction of two identical charged particles is always repulsive~\cite{Neu1999,Sader1999}.     Modifications have been made to the Poisson-Boltzmann framework to incorporate the effects of fluctuation and correlation\cite{Attard1996, Svensson1990, Attard1987, Attard1988, Podgornik1990}.  And attraction between highly-charged macroions and counterions have been demonstrated via Monte Carlo simulations~\cite{Gulbrand1984}, numericals methods of hypernetted chain integral equation~\cite{Kjellander1984, Kjellander1986} and the local-density-functional theory~\cite{Stevens1990, Diehl1999}.  Netz and Orland have formulated the problem of fluctuating counterion distribution in the presence of a fixed charge distribution in the language of field theory~\cite{Netz00a, Netz00b, Netz01a, Netz01b, Netz03}.  In particular, they have computed corrections to the electrostatic potential, counterion density, and interaction between charged particles in the presence of a fixed charge distribution in both the Poisson-Bolzmann limit and the strong coupling limit.   They have also applied field-theoretic methods to obtain the first few coefficients of a systematic low-energy expansion of the free energy for these charged systems as well as to investigate issues regarding charge renormalization and fluctuation-induced modification of the static van der Waals interactions of salt ions~\cite{Netz00a, Netz00b, Netz01a, Netz01b, Netz03}.  The problem of a charged plate in salt solution has previously been looked at by Dean and Horgan in the context of soap films~\cite{Horgan02, Horgan03, Horgan04a, Horgan04b}.   They too have formulated the problem in a field-theoretic language.  However, their results have been obtained only after approximating the action to quadratic order in all terms.  And their calculations involve mainly the total pressure exerted on the film which does not require an explicit computation of the free energy of the system.

In this paper, we extend previous works in two important ways.  We systematically compute one-loop corrections, without further approximation, to the interaction between two charged ions near the plate due to their fluctuation.  We also obtain an explicit expression for the one-loop free energy.  Similar in approach to the previous works,  we formulate our problem in a field-theoretic language that naturally incorporates the effect of fluctuation and provides a recipe for the direct computation of higher-loop corrections to various quantities.  We show that the nonlinear Poisson-Boltzmann equation can be derived as the saddle-point of the full field-theory action.  

In what follows, we will first establish the field-theoretic formalism and derive the Poisson-Boltzmann equation from first principles.  The main steps involve writing down the microscopic Hamiltonian for the single charged plate in a generally asymmetric salt solution and rewrite the resulting partition function in terms of field-theoretic variables.  This will be done in Section II of this paper.  We will then specialize to the case of symmetric salt ions and briefly summarize results from the mean-field approximation and discuss the relevant physics for these salty systems in Section III.  In the subsequent section, we will systematically consider corrections to the mean-field predictions when we incorporate the effect of fluctuating salt-ions densities .   We will first expand the action around the saddle point and, using techniques of Green's function and functional determinant, compute one-loop correction to the free energy density.  We will also consider  the interaction between ions in the presence of the charged plate and their self energies in section IV.  Finally, we summarize and conclude in Section V.
\medskip

\section{Field-theory Formulation}  

In this section, we formulate the problem of a charged plate in salt solution in a field-theoretic language and establish notations.  The field-theoretic approach has been previously established in the study of systems containing counterions and symmetric salts~\cite{Netz00a, Netz00b, Netz01a, Netz01b, Netz03, Horgan02, Horgan03, Horgan04a, Horgan04b}.  We hereby extend the formulation to the general case of assymmetric salts.  We consider a system that consists of $N_-$ point-like particles of charge $-Z_-\left\vert e\right\vert$ and $N_+$ point-like particles of charge $+Z_+\left\vert e \right\vert$, as well as a single charged surface of constant charge density.  Without loss of generality, we assume the surface to possess a negative charge density $\sigma({\bf{x}})=-\left\vert e\right\vert n_0\delta(z)=-\left\vert e\right\vert n_f({\bf{x}})$;  results for a positively charged surface can be obtained with minimal modification.  The charge neutrality condition can then be written as
\begin{eqnarray}
-ZeN_-+ZeN_+-\left\vert e\right\vert n_0A&=&0\;,
\end{eqnarray}
\\
where $A$ is the area of the plane.  Strictly speaking, the overall neutrality of the system must imply the relation of $N_+>N_-$ for the case of negatively charged surface;  both the counterions and salt cations contribute to the overall distribution of positive charges.  In what follows, we will not make the distinction between the counterions and cations; we will simply refer to both species as counter-ions while the salt anions as coions.  The energy of the system can be simply written as:
\begin{eqnarray}
\beta E_N&=&Z_-^2l_B\sum^{N_-}_{j>k}\,\frac{1}{\vert{\bf{x}}^-_j-{\bf{x}}^-_k\vert}+Z_+^2l_B\sum^{N_+}_{j>k}\,\frac{1}{\vert{\bf{x}}^+_j-{\bf{x}}^+_k\vert}\nonumber\\
&&-Z_+Z_-l_B\sum^{N_+}_{j=1}\sum^{N_-}_{k=1}\,\frac{1}{\vert{\bf{x}}^-_j-{\bf{x}}^+_k\vert}\nonumber\\
&&+Z_-\sum^{N_-}_{j}\,\phi({\bf{x}}^-_j)-Z_+\sum^{N_+}_{j}\,\phi({\bf{x}}^+_j)\;,
\label{Hamiltonian}
\end{eqnarray}
where we have defined 

\begin{eqnarray}
\phi(\bf{x}^{\pm})&=& l_B\int\,d{\bf{x'}}\frac{n_f({\bf{x'}})}{\vert{\bf{x}}^{\pm}-{\bf{x'}}\vert}
\end{eqnarray}
as the "external" field induced by the presence of the charged plate, and the Bjerrum length $l_B=e^2/\epsilon kT$ describes the length scale where electrostatic interaction dominates thermal fluctuation.  At room temperature, $l_B\approx 7$$\AA$ for water.  It is natural to define a number density operator ${\hat{\rho}}({\bf{x}}^{\pm})=\sum^{N_{\pm}}_{i=1}\,\delta({\bf{x}}^{\pm}-{\bf{x}}_i)$ for both the counterions and coions.  In terms of these density operators, we can construct the canonical partition function associated with the interaction energy  in Eq. (\ref{Hamiltonian}) as follows:
\begin{widetext}
\begin{eqnarray}
Z_{N_+,N_-}&=&\frac{1}{N_+!}\prod^{N_+}_i\left[\int \frac{d{\bf{x}}^+_i}{a^3}\right]\,\frac{1}{N_-!}\prod^{N_-}_i\left[\int \frac{d{\bf{x}}^-_i}{a^3}\right]\exp\left\{Z^2_-N_{-}V(0)+Z^2_+N_{+}V(0)-Z_-\sum^{N_-}_{j}\,\phi({\bf{x}}^-_j)+Z_+\sum^{N_+}_{j}\,\phi({\bf{x}}^+_j)\right.\nonumber\\
&&-\left.\frac 12\int d{\bf{x}}\,d{\bf{x}}'\,\left[Z_+{\hat{\rho}}({\bf{x}}^+)-Z_-{\hat{\rho}}({\bf{x}}^-)\right]V({\bf{x,x'}})\left[Z_+{\hat{\rho}}({\bf{x'}}^+)-Z_-{\hat{\rho}}({\bf{x'}}^-)\right]\right\}\;,
\label{energy}
\end{eqnarray}
\end{widetext}
where $V({\bf{x,x'}})=\frac{l_B}{\left\vert{\bf{x-x'}}\right\vert}$ is the Coulomb interaction between ions, and  as such $V(0)$ represents the Coulomb self-energy.  In the form written in Eq. (\ref{energy}), the Coulomb interaction acts on the effective charge density ${\hat{\rho}}_c=Z_+{\hat{\rho}}({\bf{x}}^+)-Z_-{\hat{\rho}}({\bf{x}}^-)$ which is physically modulated by the spacial distribution of each ion species and their respective valence.  From now on, we will assume that the total number of ions (counter-ions and salt) in solution is conserved, i.e. $N_++N_-=N$. 

To elevate Eq. (\ref{energy}) to a partition function of field variables, we borrow techniques established in field theory and make use of the following resolution of the identity operator 
\begin{widetext}
\begin{eqnarray}
\hat{\bf{1}}&=&\int{\mathcal{D}}\rho^{\pm}\,\delta\left(\rho^{\pm}-{\hat{\rho}}({\bf{x}}^{\pm})\right)=\int{\mathcal{D}}\rho^{\pm}\,{\mathcal{D}}\psi^{\pm}\,\exp\left\{i\,\int d{\bf{x}}\,\psi^{\pm}({\bf{x}})\left[\rho^{\pm}({\bf{x}})-{\hat{\rho}}({\bf{x}}^{\pm})\right]\right\}\;.
\label{identity}
\end{eqnarray}
\end{widetext}
Upon insertion of Eq. (\ref{identity}) into Eq. (\ref{energy}) and simplifying, we can now rewrite our canonical partition function in terms of field variables $\rho^{\pm}({\bf{x}})$ and $\psi^{\pm}({\bf{x}})$:
\begin{widetext}
\begin{eqnarray}
Z_{N_+,N_-}&=&\int{\mathcal{D}}\rho^+{\mathcal{D}}\rho^-{\mathcal{D}}\psi^+{\mathcal{D}}\psi^-\,\exp\left\{-\frac 12\int d{\bf{x}}\,d{\bf{x}}'\,\left[Z_+\rho^{+}({\bf{x}})-Z_-\rho^-({\bf{x}})\right]V({\bf{x,x'}})\left[Z_+\rho^{+}({\bf{x'}})-Z_-\rho^-({\bf{x'}})\right]\right.\nonumber\\
&&\left.+i\int d{\bf{x}}\left[\psi^+({\bf{x}})\rho^+({\bf{x}})+\psi^-({\bf{x}})\rho^-({\bf{x}})\right]\right\}\nonumber\\
&&\times\frac{1}{N_+!}\left[\int \frac{d{\bf{x}}^+}{a^3}\,e^{Z^2_+V(0)+Z_+\phi({\bf{x}}^+)-i\,\psi^+({\bf{x}}^+)}\right]^{N_+}\,\frac{1}{N_-!}\left[\int \frac{d{\bf{x}}^-}{a^3}\,e^{Z^2_-V(0)-Z_-\phi({\bf{x}}^-)-i\,\psi^-({\bf{x}}^-)}\right]^{N_-}\;.\nonumber\\
\label{fieldpartition}
\end{eqnarray}
\end{widetext}
In the form written in Eq. (\ref{fieldpartition}), the field-theoretic version of the canonical partition function $Z_{N_+,N_-}[\rho^{\pm},\psi^{\pm},\phi]$ is rather cumbersome.  To bring the expression to a more manageable form, we will introduce a chemical potential (in unit of $k_BT$) and consider the grand canonical ensemble:
\begin{eqnarray}
{\mathcal{Z}}_{\mu_+,\mu_-}&=&\sum_{N_-,N_+=0}^{\infty}\left(e^{\mu_+}\right)^{N_+}\left(e^{\mu_-}\right)^{N_-}\,Z_{N_+,N_-}\;.
\label{grand}
\end{eqnarray}
We also explicitly perform the functional integral over field variables $\rho^{\pm}({\bf{x}})$ in Eq. (\ref{fieldpartition}).  The operation involves performing a functional Gaussian integration followed by another integration over a delta functional which imposes the condition that $\frac{\psi^+}{Z_+}=-\frac{\psi^-}{Z_-}$.  Defining a new field $\Psi=\frac {1}{2}\left(\frac{\psi^+}{Z_+}-\frac{\psi^-}{Z_-}\right)$, we can explicitly write the grand-canonical partition function in Eq. (\ref{grand}) as follows:  
\begin{eqnarray}
{\mathcal{Z}}_{\mu_+,\mu_-}&=&\mathcal{N}_0\,\int{\mathcal{D}}\Psi\,e^{-{\mathcal{S}}\left[\Psi,\phi\right]}\;,
\end{eqnarray}
where $\mathcal{N}_0$ is the usual normalization constant and 
\begin{widetext}
\begin{eqnarray}
{\mathcal{S}}\left[\Psi,\phi\right]&=&\frac{1}{4\pi l_B}\int d{\bf{x}}\,\left[\frac12\Psi({\bf{x}})\left(-\nabla^2\right)\Psi({\bf{x}})-\frac{4\pi l_B}{a^3}e^{Z^2_+V(0)+Z_+\phi({\bf{x}})+i\,Z_+\Psi({\bf{x}})+\mu_+}-\frac{4\pi l_B}{a^3}e^{Z^2_-V(0)-Z_-\phi({\bf{x}})-i\,Z_-\Psi({\bf{x}})+\mu_-}\right]\;,\nonumber\\
\label{8}
\end{eqnarray}
\end{widetext}
where we have substituted the explicit expression for the inverse of the Coulomb operator $4\pi V^{-1}({\bf{x,x'}})=-\nabla^2\delta({\bf{x-x'}})$.  We can then bring the grand-canonical partition function  in Eq. (\ref{8}) into the standard form
\begin{widetext}
\begin{eqnarray}
{\mathcal{Z}}_{\mu_+,\mu_-}&=&\int{\mathcal{D}}\Psi\,e^{-{\mathcal{S}}\left[\Psi,\phi\right]}\;,\nonumber\\
\mathcal{S}[\Psi,\phi]&=&\frac{1}{\ell_B}\int d{\bf{x}}\,\left[\frac 12\,\Psi({\bf{x}})\left(-\nabla^2\right)\Psi({\bf{x}})-\frac{\kappa_+^2}{2}\,e^{Z_+\left[i\Psi({\bf{x}})+\phi({\bf{x}})\right]}
-\frac{\kappa_-^2}{2}\,e^{-Z_-\left[i\Psi({\bf{x}})+\phi({\bf{x}})\right]}\right]\;,
\label{action}
\end{eqnarray}
\end{widetext}
by defining the quantity $\ell_B$, which has the dimension of length and is simply a rescaling of the Bjerrum length;  the quantities $\kappa_{\pm}$ which will play the role of inverse screening length of the counterions and coions, respectively; and the quantities $\rho^{\pm}_0$  which can be interpreted as the bulk densities of the counterions and coions, respectively, in solution:
\begin{eqnarray}
\ell_B&=&4\pi\,l_B\;,\\
\kappa_{\pm}^2&=&2\rho_0^{\pm}\ell_B\;,\\
\rho^{\pm}_0&=&\frac{e^{Z_{\pm}^2V(0)+\mu_{\pm}}}{a^3}\;.
\end{eqnarray}
Let us define $\phi_{\pm}({\bf{x}})=Z_{\pm}\phi({\bf{x}})$.  Based on our construction of the partition function, the average counterion and coion densities $\langle\rho_{\pm}({\bf{x}})\rangle$ can be readily related to the grand-canonical partition function:
\begin{eqnarray}
\langle\rho_{\pm}({\bf{x}})\rangle&=&\frac{\delta\,\log Z_{\mu_+,\mu_-}[\phi]}{\delta\phi_{\pm}({\bf{x}})}\;.
\label{density}
\end{eqnarray}
Extension to higher-order functional derivatives of the partition function with respect to the external field $\phi({\bf{x}})$ brings about higher-order cumulant correlation functions in the standard way.  From this point on, we will limit ourselves to consider only the case where both the counterions and coions have the same valency $Z_{\pm}=Z$ and the same chemical potential $\mu_{\pm}=\mu$.   We note that the chemical potential in the grand canonical potential is not uniquely defined, since the transformation $\Psi({\bf{x}})\rightarrow\Psi({\bf{x}})+i\,\mu_0$ and $\mu\rightarrow\mu-\mu_0$ leaves the free energy invariant.  We also note that the ``external" field $\phi({\bf{x}})$, determined by 
\begin{equation}
-\nabla^2\phi({\bf{x}})=4\pi\,l_Bn_f({\bf{x}})\;,
\end{equation}
contains an arbitrary constant.  To make connection with the standard Poisson-Boltzmann equation, we determine the saddle point for the action in Eq. (\ref{action}), with the conditions that $\kappa_{\pm}=\kappa_0$ and $Z_{\pm}=Z$, via the Euler-Lagrange equation:
\begin{eqnarray}
\nabla^2[i\Psi_0({\bf{x}})]&=&\frac{Z\kappa_0^2}{2}\,e^{Z\left[i\Psi_0({\bf{x}})+\phi({\bf{x}})\right]}\nonumber\\
&&-\frac{Z\kappa_0^2}{2}\,e^{-Z\left[i\Psi_0({\bf{x}})+\phi({\bf{x}})\right]}\;.
\label{PBfore}
\end{eqnarray}
To bring Eq. (\ref{PBfore}) into a more familiar form, we define a shift in the field variable: $\varphi({\bf{x}})=Z\left[i\Psi_0({\bf{x}})+\phi({\bf{x}})\right]$.  In terms of the new field $\varphi$, Eq. (\ref{PBfore}) becomes
\begin{eqnarray}
\nabla^2[\varphi({\bf{x}})]-\kappa^2\,\sinh[\varphi({\bf{x}})]&=&-4\pi Z\,l_Bn_f({\bf{x}})\;,
\label{inhomo}
\end{eqnarray}
where $\kappa^2=\kappa_0^2Z^2$.  Eq. (\ref{inhomo}) is the Poisson-Boltzmann equation for a charged plate in salt solution that contains both counterions and coions.   Our reinterpretation of the Poisson-Boltzmann equation as a saddle-point of an underlying field-theoretic action affords us a systematic way to include fluctuation effects and go beyond mean-field theory.
\begin{figure}[htb]
\vspace{2mm}
\hspace{23mm}
\includegraphics{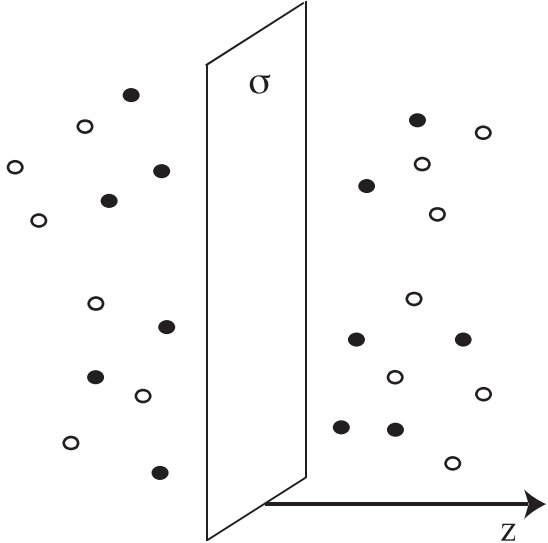}
\vspace{2mm}
\caption{A single charged plate in salt solution.  The system is symmetric with respect to $z=0$.  Without loss of generality, the charged plate is assumed to have a positive surface-charge density $\sigma$, with counterions (solid circles) and coions (open circles) present on both sides of the plate.}
\label{oneplate}
\end{figure}
\section{Mean-field Theory}

As stated, closed-form solutions exist for the Poisson-Boltzmann equation only for the planar geometry~\cite{Guoy1910, Chapman1913, Stern1924} and cylindrical geometry~\cite{VerweyText, Fuoss1951}.  In cases of other geometries, one has to resort to further approximations.  For the case of spheres, one can apply the Derjaguin approximation when the interaction potentials between spheres are much larger than the range of the potential~\cite{VerweyText}.  Under this condition one may simply neglect curvature effects at the lowest order and adapt results from the case of flat plates.  Thus, there is an intrinsic usefulness in understanding the solution of the Poisson-Boltzmann equation for even the simple case of planar geometry.

The solution of the full nonlinear Poisson-Boltzmann equation for a single charged plate in salt solution is well known in literature.  We will provide here a brief summary of results and the relevant physics of the system.  First, we note that the single charged-plate system is a one-dimensional problem whose solution for $0\leq z\leq \infty$ is related to solution for  $-\infty\leq z\leq 0$ by symmetry  (Fig. \ref{oneplate}).  To solve Eq. (\ref{inhomo}) with the delta-function source term,  we simply consider the associated homogeneous equation and match the following boundary conditions imposed by the charged surface:
\begin{eqnarray}
\label{infinitycondition}
\left.\frac{d\varphi_0(z)}{dz}\right\vert_{z\rightarrow\infty}&=&0\\
\left.\frac{d\varphi_0(z)}{dz}\right\vert_{z\rightarrow 0}&=&\frac{Z\sigma_s\,\ell_B}{2e}\equiv \frac{2}{\lambda}
\label{charge}
\end{eqnarray}
where $\lambda=\frac{4e}{Z\sigma_s\ell_B}$ is the Gouy-Chapman length.  The boundary condition at the plate in Eq. (\ref{charge}) is found by integrating the Poisson-Boltzmann equation (\ref{inhomo}) over the interval $-\epsilon< z < \epsilon$ and taking the limit as $\epsilon\rightarrow 0$.  We have also introduced a useful length scale $\lambda$ in Eq. (\ref{charge}) for future convenience.  The solution for the mean-field electrostatic potential $\varphi_0(z)$ can be readily found:
\begin{eqnarray}
\varphi_0(z)&=&-4\,\tanh^{-1}\,e^{-\kappa(\vert z\vert+z_0)}\;,
\label{MFsoln}
\end{eqnarray}
where the constant of integration $z_0$ is completely determined by the relation
\begin{eqnarray}
\kappa\lambda&=&\sinh\,\kappa z_0\;.
\end{eqnarray}
For this problem, we can construct two dimensionless parameters that completely characterize the system.  The first useful dimensionless quantity $\frac{\ell_B}{\lambda}$ encodes information about the charge density of our plate.  Given a highly charged plate in salt solution, $\frac{\ell_B}{\lambda}\gg 1$.  The second dimensionless quantity $\kappa\lambda$, on the other hand, yields information on the charge densities of the salt ions.  From Eq. (\ref{MFsoln}), we see that $\kappa^{-1}$ represents the screening length of the ions.  Since a highly charged background of ions implies stronger screening, the condition that $\kappa\lambda\gg 1$ indicates either a dense distribution of moderately charged salt ions or high valency from the ions themselves.  Conversely the limit of $\kappa\lambda\ll 1$ corresponds to the case when only counterions are present.  In what follows, we will see that these two dimensionless combinations of physical parameters also provide a convenient choice when expressing some of our more complicated results, in a way that would highlight their underlying physics.  Meanwhile using the mean-field electrostatic potential $\varphi_0$ given by Eq. (\ref{MFsoln}), we can compute the corresponding charge density for the ions as well as the free-energy density at the mean-field level.  First, the counterion and coion charge densities, as a function of distance $z$ from the charged plate,  can now be written down in a straightforward manner:
\begin{eqnarray}
\label{rho00}
\rho_{\pm}^{(0)}(z)&=&\rho_0\,e^{\,\pm\,\varphi(z)}\;,\\
&=&\rho_0\left(\frac{1\pm e^{-\kappa(z+z_0)}}{1\mp e^{-\kappa(z+z_0)}}\right)^2\;.
\label{rho0}
\end{eqnarray}
\\
Our definition for the salt-ions charge densities in Eq. (\ref{rho00}) is consistent with the form of $\varphi_0(z)$ with the convention of a plus sign.  The Helmholtz free-energy density is given by
\begin{eqnarray}
\beta F&=& \int d{\bf{x}}\,\rho_-({\bf{x}})\left[\log\,\rho_-({\bf{x}})\,a^3-1\right]\nonumber\\
&&+\int d{\bf{x}}\,\rho_+({\bf{x}})\left[\log\,\rho_+({\bf{x}})\,a^3-1\right]\nonumber\\
&&+\frac{1}{2\ell_B}\int d{\bf{x}}\left[\nabla\varphi({\bf{x}})\right]^2\;,
\label{bF}
\end{eqnarray}
\\
where $a$ is the typical ionic size of the charges.  The first two terms are the entropy of the counter- and co-ions, while the last term is the electrostatic energy of the system.  The free energy $\beta F$ can in general be divided into a volume contribution $f_V$ and an area contribution $f_A$:  $\beta F=f_vV+f_AA$.  Upon substitution of Eqs. (\ref{MFsoln}) and (\ref{rho0}) into Eq. (\ref{bF}) and further evaluation, we can find the mean-field free-energy densities $f_V$ and $f_A$ as a function of $x=\kappa\lambda$:
\begin{widetext}
\begin{eqnarray}
\label{FA}
\beta f_V&=&\rho_0\left(\log\,\rho_0\,a^3-1\right)\\
\beta f_A&=&-\frac{2}{\ell_B\lambda\left(x+\sqrt{x^2+1}\right)}-\frac{2}{\ell_B\lambda}\left(x-\sqrt{1+x^2}\right)\log\left( \frac{x^2}{4}\right)-\frac{2}{\ell_B\lambda}\left(x-\sqrt{1+x^2}\right)\log \left(\frac{2a^3}{\ell_B\lambda^2}\right)\nonumber\\
&&-\frac{2}{\ell_B\lambda}\sqrt{1+x^2}\log\left(\frac{\sqrt{x^2+1}-1}{\sqrt{x^2+1}+1}\right)\;.
\label{F00}
\end{eqnarray}
\end{widetext}
The plot of Eq. (\ref{F00}) is shown in Fig. ({\ref{F000}).  For simplicity, we assume that $a^3\sim\ell_B\lambda$ and that the free-energy density is units of $\ell_B\lambda$.  In the limit that $\kappa\rightarrow 0$, the total free energy of the system based on Eqs. (\ref{FA}) and (\ref{F00}) correctly reduce to that for a system with counterions only, which has the form of the free energy of an ideal gas.
\begin{figure}[htb]
\vspace{2mm}
\hspace{23mm}
\includegraphics{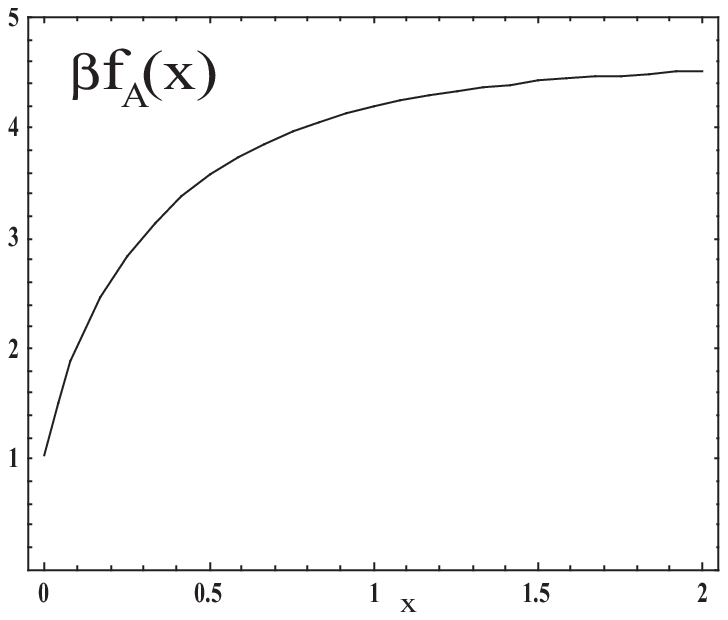}
\vspace{2mm}
\caption{Plot of the effective mean-field free-energy density as a function of $x=\kappa\lambda$.  The free-energy density is measured in units of $\ell_B\lambda$.  We assume that $a^3\sim\ell_B\lambda^2$.}
\label{F000}
\end{figure}

\section{Fluctuation effects}

With the field-theoretic formulation in place for a single charged surface in salt solution, we can now systematically investigate effects of fluctuating salt-ion densities.   In what follows, we will first describe the general methodology of our calculations before presenting details of our results in following subsections.  To go beyond mean-field, we first expand the action in Eq. (\ref{8}) about the saddle point $\Psi_0$:
\begin{widetext}
\begin{eqnarray}
{\mathcal{S}}[\Psi,\phi]&=&{\mathcal{S}}[\Psi_0,\phi]+\int d^3{\bf{x}}\,\left.\frac{\delta{\mathcal{H}}}{\delta\Psi({\bf{x}})}\right\vert_{\Psi=\Psi_0}\Delta\Psi({\bf{x}})+\frac12\int d^3{\bf{x}}\int d^3{\bf{y}}\left.\frac{\delta^2{\mathcal{H}}}{\delta\Psi({\bf{x}})\delta\Psi({\bf{y}})}\right\vert_{\Psi=\Psi_0}\Delta\Psi({\bf{x}})\Delta\Psi({\bf{y}})+\cdots\;,
\label{action2}
\end{eqnarray}
where ${\mathcal{H}}$ is the Hamiltonian given by
\begin{eqnarray}
\mathcal{H}&=&\frac{1}{\ell_B}\left\{\frac 12\,\Psi({\bf{x}})\left(-\nabla^2\right)\Psi({\bf{x}})-\kappa^2\cosh\left[Z\left[i\Psi({\bf{x}})+\phi({\bf{x}})\right]\right]\right\}\;,
\end{eqnarray}
\end{widetext}
for the case of symmetric salt ions and $\Delta\Psi({\bf{x}})=\Psi({\bf{x}})-\Psi_0({\bf{x}})$.  By definition, the saddle point satisfies the mean-field equation, and the linear term in Eq. (\ref{action2}) vanishes accordingly.  Since the quadratic term is Gaussian in nature, we can formally write the partition function of this quadratic action as a functional determinant~\cite{PeskinText}.  To compute this determinant, we define the kernel ${{\hat{K}}({\bf{x,y})}}=\frac{\delta^2{\mathcal{H}}}{\delta\psi({\bf{x}})\delta\psi({\bf{y}})}$, and we observe that its operator inverse is just the Green's function $G({\bf{x,y}})$ defined by:
\begin{eqnarray}
\int d{\bf{y}} \;{{\hat{K}}({\bf{x,y})}}\,G({\bf{y,x'}})=\delta({\bf{x-x'}})\;.
\label{KG}
\end{eqnarray}
The one-loop contribution to the free-energy density, representing the change in free energy due to fluctuations of the salt ions, can then be formally written as
\begin{eqnarray}
\beta\Delta F&=&\frac 12 \log\det {\hat{K}}-\frac 12\log\det \left[-{\bf{{\hat{\nabla^2_x}}}}\right]\;.
\label{deltaF}
\end{eqnarray}
To evaluate $\beta\Delta F$ explicitly, we will employ the following mathematical identity for a general operator $\hat{X}$:
\begin{eqnarray}
\delta\log\det {\hat{X}}&=&{\rm{Tr}}\,{\hat{X}}^{-1}\delta{\hat{X}}\;.
\label{det}
\end{eqnarray}
We will first differentiate the free-energy density in Eq. (\ref{deltaF}) with respect to $\ell_B$.  Using Eq. (\ref{det}),  we can express the functional determinant of the kernel ${{\hat{K}}({\bf{x,y})}}$ in terms of its operator inverse, the Green's function $G({\bf{x,y}})$.  In the process, we need to make provisions to subtract out the (infinite) self-energy in the resulting expressions.  In the form written in Eq. (\ref{KG}), we can regard the Green's function $G({\bf{x,y}})$ as a fluctuating potential at point ${\bf{x}}$ generated by a test charge around which there exist fluctuating charges.  Thus, the Green's function contains information about correlations of the system, and in particular, can be interpreted as the electrostatic interaction between two test charges located at ${\bf{x}}$ and ${\bf{y}}$ in the presence of fluctuating ions in the bulk.  We will now present the details on the computation of the Green's function.

\subsection{Construction of the Green's function}

To determine the Green's function, we use our action in Eq. (\ref{action}) to compute the kernel ${{\hat{K}}({\bf{x,y})}}$, and we can write the representation of Eq. (\ref{KG}) in real space:
\begin{eqnarray}
\left\{-\nabla^2+\kappa^2{\rm{cosh}}\left[\varphi_0\left(z\right)\right]\right\}\,G({\bf{x,x'}})=\ell_B\,\delta(\bf{x-x'})\;.
\label{GG}
\end{eqnarray}
We note that the Green's function $G({\bf{x,x}}')$ contains lateral invariance and can thus be Fourier-transformed in the directions perpendicular to the surface.  Upon substituting the saddle-point solution $\varphi_0(z)$ [Eq. (\ref{MFsoln})] into the kernel, we obtain 
\begin{eqnarray}
\left[-\frac{\partial^2}{\partial u^2}+q^2+\kappa^2+2\kappa^2\,{\rm{csch}}^2\kappa \left(z+z_0)\right)\right]\,G(z,z';q)&&\nonumber\\
=\ell_B\delta(z-z'),&&\nonumber\\
\label{u-equation}
\end{eqnarray}
where $q^2=q_x^2+q_y^2$.  In order to solve Eq. (\ref{u-equation}), we have to obtain the solution to the corresponding homogeneous equation:
\begin{eqnarray}
\left[-\frac{\partial^2}{\partial (\kappa\,u)^2}+4\alpha^2+2\,{\rm{csch}}^2\,\kappa u\right]\,h(u)&=&0\;,
\label{h}
\end{eqnarray}
where we have made a shift of variable $u=z+z_0$ and labelled the homogeneous solution $h(u)$ and defined the quantity $\alpha^2=\frac14\left(1+\frac{q^2}{\kappa^2}\right)$.  The two independent homogeneous solutions are:
 \begin{eqnarray}
\label{elementarysolution1}
h_1(u)&=&e^{-2\alpha \kappa u}\left(1+\frac{{\rm{coth}}\,\kappa u}{2\alpha}\right)\;,\\
h_2(u)&=&e^{2\alpha \kappa u}\left(1-\frac{{\rm{coth}}\,\kappa u}{2\alpha}\right)\;.
\label{elementarysolution2}
\end{eqnarray}
To construct the Green's function $G(z,z')$, we split space into three distinct regions and define in each of the region (Fig.\,\ref{GreensFunction}):
\begin{figure}[htb]
\vspace{2mm}
\hspace{23mm}
\includegraphics{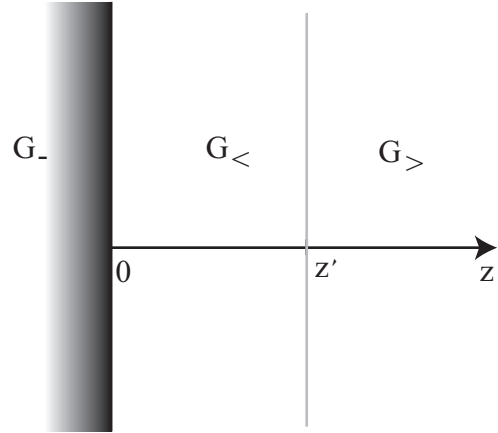}
\vspace{2mm}
\caption{Definition of the Green's function in three different regions:  $z <\infty$, $0<z<z'$, and $z>z'$.  See Eqs. (\ref{a}), (\ref{b}), (\ref{c}).}
\label{GreensFunction}
\end{figure}
\begin{eqnarray}
\label{a}
G_{>}(z,z')&=&A(z')h_1(z),\,\,\,\,\,\,\,\,\,\,\,\,\,\,\,\,\,\,\,\,\,\,\,\,\,\,\,\,\,\,\,\,\,\,\,\,\,\,\,\,z>z'\\
G_{<}(z,z')&=&B(z')h_1(z)+C(z')h_2(z)\;,\nonumber\\
\label{b}
&&\,\,\,\,\,\,\,\,\,\,\,\,\,\,\,\,\,\,\,\,\,\,\,\,\,\,\,\,\,\,\,\,\,\,\,\,\,\,\,\,\,\,\,\,\,\,\,\,\,\,\,\,\,\,\,\,\,\,\,\,0<z<z'\\
G_{-}(z,z')&=&D(z')h_1(-z)\;,\,\,\,\,\,\,\,\,\,\,\,\,\,\,-\infty<z<0
\label{c}
\end{eqnarray}
To determine the coefficients $A(z'), B(z'), C(z')$, and $D(z')$, we impose the following boundary conditions:
\begin{eqnarray}
\label{A1}
G_-(0,z')&=&G_<(0,z')\\
\label{A2}
\frac{\partial G_-}{\partial z}\vert_{z=0}&=&\frac{\partial G_<}{\partial z}\vert_{z=0}\\
\label{B1}
G_<(z,z')\vert_{z=z'}&=&G_>(z,z')\vert_{z=z'}\\
\frac{\partial G_<}{\partial z}\vert_{z=z'}-\frac{\partial G_>}{\partial z}\vert_{z=z'}&=&\ell_B
\label{B2}
\end{eqnarray}
The first three conditions in Eqs. (\ref{A1}), (\ref{A2}), and ({\ref{B1}}) simply reinforce the continuity of the Green's function $G(z,z')$ across the boundary $z=0$.  The presence of a charged-surface boundary at $z=0$ has already been incorporated in the form of the mean-field solution $\varphi_0(z)$ [Eq.(\ref{MFsoln})], which we used in obtaining Eq. (\ref{h}).  Therefore, the Green's function, which represents the correlation between two charges at distances $z$ and $z'$ from the charged surface, along with its derivative should be well-behaved and not contain any singularity.  The last condition in Eq. (\ref{B2}) represents the location of a test charge at $z'$ as encoded by the delta function in Eq. (\ref{GG}).  So far in specifying the boundary conditions in Eqs. (\ref{A1})-(\ref{B2}) we have only considered the case when $z'>0$;  results for the case of $z'<0$ can be deduced from symmetry.  After some algebraic computation, we arrive at the following expressions for the Green's function $G(z,z')$ for the region $0<z<z'$:
\begin{widetext}
\begin{eqnarray}
G(z,z';q)&=&\frac{\ell_B}{2q^2}\sqrt{\kappa^2+q^2}\left[e^{-2\alpha\kappa(z'-z)}\left(1+\frac{{\rm{coth}}\,\kappa(z'+z_0)}{2\alpha}\right)\,\left(1-\frac{{\rm{coth}}\,\kappa(z+z_0)}{2\alpha}\right)\right.\nonumber\\
&&+\left.{\mathcal{M}}(q)\,e^{-2\alpha\kappa(z'+z)}\left(1+\frac{{\rm{coth}}\,\kappa(z'+z_0)}{2\alpha}\right)\,\left(1+\frac{{\rm{coth}}\,\kappa(z+z_0)}{2\alpha}\right)\right]\\
\nonumber\\
\label{Green}
\end{eqnarray}
where
\begin{eqnarray}
{\mathcal{M}}(q)&=&\frac{\sqrt{1+(\kappa\lambda)^2}}{\left[\sqrt{(q\lambda)^2+(\kappa\lambda)^2}+\sqrt{1+(\kappa\lambda)^2)}\right]\left[1+\sqrt{\left((q\lambda)^2+(\kappa\lambda)^2\right)\left(1+(\kappa\lambda)^2\right)}+(\kappa\lambda)^2+(q\lambda)^2\right]}\;.
\label{M}
\end{eqnarray}
\end{widetext}
The Green's function for the other regions ($G_>$ and $G_-$) can also be computed in a straightforward manner.  For our purposes, we are only interested in the Green's function in the case of $z=z'$, which represents self-correlation.  $G(z,z;q)$ takes the following expression:
\begin{widetext}
\begin{eqnarray}
G(z,z;q)&=&\frac{\ell_B}{2q^2}\sqrt{\kappa^2+q^2}\left[1-\frac{{\rm{coth}}^2\,\kappa(z+z_0)}{4\alpha^2}+{\mathcal{M}}(q)e^{-4\alpha\kappa z}\,\left(1+\frac{{\rm{coth}}\,\kappa(z+z_0)}{2\alpha}\right)^2\right]\;.\nonumber\\
\label{selfenergy}
\end{eqnarray}
\end{widetext}
To understand the expression of the correlation self-energy, we see that in the $\kappa\rightarrow 0$ limit, the ${\mathcal{M}}(q)$-factor in Eq. (\ref{selfenergy}) reduces to the expression corresponding to the case of a single charged plate with only counterions present in the system~\cite{LauThesis, Lau2002}.  One might expect that the $\kappa\rightarrow 0$ limit should correspond to a system with no ions.  However,  the condition of overall neutrality in the system necessitates the presence of counter-ions even when there is no salt present in the system. 

The self interaction is defined as the partially Fourier-transformed Green's function in Eq. (\ref{selfenergy}):
\begin{eqnarray}
G_s(z,z)&=&\int\frac{d^2q}{(2\pi)^2}\,\left[G(z,z;q)-\frac{\ell_B}{2q}\right]\;.
\end{eqnarray}
where we have subtracted out the infinite Coulomb self-energy at the end.  The plot of a rescaled $G_s(z)$ is shown in Fig.~\ref{p1}.  As apparent in the figure, the self-energy is negative and indicates a decrease in the local free energy of the particles due to the screening of salt ions.  This extends the same conclusion previously found in systems containing only counterions~\cite{Netz00b}.
\begin{figure}[htb]
\vspace{2mm}
\hspace{23mm}
\includegraphics{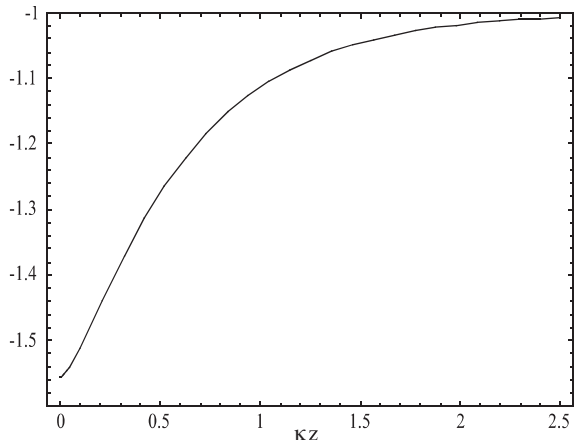}
\vspace{2mm}
\caption{Plot of the rescaled partially transformed Green's function $\frac{4\pi}{\kappa\ell_B}G_s(z, z)$ as a function of $\kappa z$ at $x=\kappa\lambda=1$.}
\label{p1}
\end{figure}

\subsection{Interaction between Two Test Charges above the Surface}

As mentioned previously, the Green's function $G({\bf{x}},{\bf{y}})$ can be interpreted as the electrostatic interaction between two test charges located at ${\bf{x}}$ and ${\bf{y}}$ in the presence of fluctuating ions in the bulk.  Using the expression of the Green's function in Eq. (\ref{selfenergy}), we can revert back to physical space in the $x$- and $y$-directions and consider the interaction of two point charges separated by a distance $r$ from each other, and both at a distance $z$ from the charged plate.  The real-space correlation self-energy relates to Eq. (\ref{selfenergy}) via a Hankel transform:
\begin{eqnarray}
G(z,r)&=&\int\frac{dq\, q}{2\pi}\,G(z,z,q)\,J_0(qr)\;,
\label{hankel}
\end{eqnarray}
where $J_0(qr)$ is the Bessel function of the first kind of order zero.  When the two charged particles are far away from the charged surface ($z\rightarrow\infty$),  the interaction can be shown to reduce analytically to the Yukawa form:
\begin{eqnarray}
G(r)&\simeq&\frac{\ell_B}{4\pi r}e^{-\kappa r}\;.
\label{Yukawa}
\end{eqnarray}
Thus as one would expect, at far enough distance the two point charges act as if though there were no charged plate at all.  On the other hand, for the two point charges that are close to the charged surface (small $z$), we can numerically integrate Eq. (\ref{hankel}) and plot the form of resulting interaction for various values of $z$, as shown in Fig. ({\ref{Greenss}).

In Fig. (\ref{Greenss}), we compare the interaction between two point charges located in close proximity to the charged plate with the usual Yukawa-type interaction experienced by the same charges in the absence of or far away from the plate.  As the distance $z$ from the charged plate decreases, we observe that the corresponding repulsion between the two point charges of the same sign is dramatically suppressed due to density fluctuation.  Therefore, we conjecture that when such other factors as Van der Waals interaction are taken into account, this decrease in electrostatic repulsion may lead to overall qualitatively different behavior for the two point-charges, such as attraction between like charges (charge inversion), with an appropriate range of system parameters and in the region close enough to the charged plate.  
\begin{figure}[htb]
\vspace{2mm}
\hspace{23mm}
\includegraphics{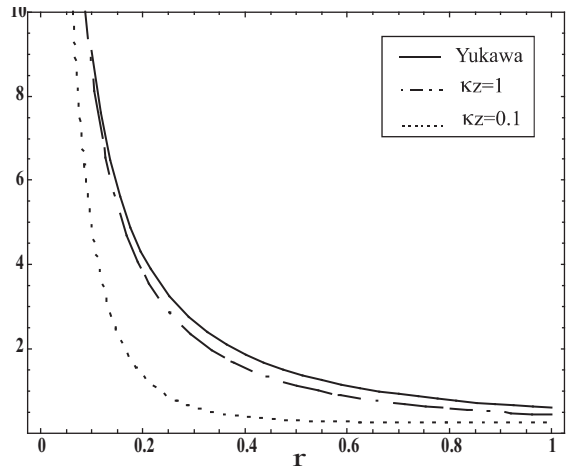}
\vspace{2mm}
\caption{Plot of the rescaled correlation self-energy $\frac{4\pi}{\kappa\ell_B}G(r,\kappa z)$ as a function of the scaled distance $\kappa z$ from the charged plate.  We use $x=\kappa\lambda=0.1$.  We see that density fluctuation of the salt ions reduces the repulsion experienced between two point charges near the plate.}
\label{Greenss}
\end{figure}

\subsection{Fluctuation Contributions to Free-Energy Density}

Armed with the Green's function in Eq. (\ref{selfenergy}), we can now calculate the fluctuation-induced correction to the free-energy density, formally expressed in terms of the functional determinant of the kernel ${\hat{K}}$:
\begin{eqnarray}
\beta\Delta F&=&\frac 12 \log\det {\hat{K}}-\frac 12\log\det \left[-{\bf{{\hat{\nabla^2_x}}}}\right]
\end{eqnarray}
The term proportional to $\det \left[-{\bf{{\hat{\nabla^2_x}}}}\right]$ comes from the normalization factor in the partition function and simply subtract out the divergence associated with the self-energy of charged particles;  a problem well-known even in the context of classical electrostatics.  Physically, analogous to the mean-field result, the one-loop free energy consists of a volume contribution $\beta\Delta f_V$ and an area contribution $\beta\Delta f_A$.  For the kernel $\hat{K}$ in Eq. (\ref{u-equation}), we see that all terms except for the last are translation-invariant.  Therefore, when integrated over all space, these translation-invariant terms would produce factors proportional to the overall volume of space and would exactly constitute the volume contribution to the one-loop free energy.   The remaining contribution has a dependence on the distance $z$ from the plate and thus constitute the area contribution to the one-loop free energy. 

Therefore, to compute the one-loop correction to the free energy due to the presence of a charged plate, we must first isolate this translation-invariant piece of the kernel from the rest that actually depends on the ions' distance from the charged plate.   This translationally invariant piece in the kernel ${\hat{K}}$ corresponds precisely to what one gets in the Debye-H\"{u}ckel approximation for the salt case, where the electrostatic potential $\varphi$ is linearized:
\begin{eqnarray}
{\hat{K}}_{DH}&=&\left(-\nabla^2_{\bf{x}}+\kappa^2\right)\delta({\bf{x-x'}})\;.
\end{eqnarray}
The corresponding Green's function for the Debye-H{\"{u}}ckel kernel is simply the Yukawa potential:
\begin{eqnarray}
G_{DH}(z,z';q)&=&\frac{\ell_B}{2\sqrt{\kappa^2+q^2}}\,e^{-2\alpha\kappa\vert z-z' \vert}\;.
\label{DH}
\end{eqnarray}
which shows explicit translational invariance.  To determine this volume contribution to the one-loop free energy, we can simply consider the change in the internal energy density when a particle of charge $e$ is brought into an electrolyte solution;  the modification of this self-energy is exactly given by half the difference between the Yukawa potential and Coulomb potential in the limit of zero separation~\cite{LauThesis}.  Using standard thermodynamic relation between the internal energy and the free energy and the condition that $\Delta f_V\rightarrow 0$ as $T\rightarrow\infty$, we readily obtain the following result:
\begin{eqnarray}
\beta\Delta f_VV&=&\frac 12 \log\det {\hat{K}}_{KH}-\frac 12\log\det \left[-{\bf{{\hat{\nabla^2_x}}}}\right]\nonumber\\
&=&-\frac{\kappa^3}{12\pi}V\;.
\end{eqnarray}
This is the classical result obtained by Debye and H{\"{u}}ckel~\cite{DH}.  Thus, subtracting off the volume contribution, we have the following expression for $\beta\Delta f_A$:
\begin{eqnarray}
\beta\Delta f_AA &=&\frac 12 \log\det {\hat{K}}-\frac 12 \log\det {\hat{K}}_{DH}\;.
\label{deltaF2}
\end{eqnarray}
To proceed, we differentiate Eq. (\ref{deltaF}) with respect to $\ell_B$ and apply the identity in Eq. (\ref{det}).  As a result, the interesting part of the one-loop correction to the free-energy density is
\begin{widetext}
\begin{eqnarray}
\frac{\partial\beta\Delta f_A}{\partial\ell_B}&=&\frac{1}{\ell_B}\left\{\int\frac{d^2q}{(2\pi)^2}\int^{\infty}_0 dz\left[G(z,z;q)\right]\frac{\partial}{\partial\ell_B}\left[\kappa^2\cosh\varphi_0(z)\right]-\int\frac{d^2q}{(2\pi)^2}\int^{\infty}_0 dz\left[G_{DH}(z,z;q)\right]\,\frac{\partial}{\partial\ell_B}\left[\kappa^2\right]+\ldots\right\}\;,\nonumber\\
\label{oneloop}
\end{eqnarray}
\end{widetext}
where $A$ has the dimension of area.  At this point, we need to make one more adjustment to render the expression finite:  we need to subtract off by the part in the Green's function that corresponds to the free-space case where there are no ions present, $G_0(q)=\ell_B/2\left\vert q\right\vert$, from each of the Green's functions in Eq. (\ref{oneloop}).  Upon further simplification, we can express the one-loop correction to the free-energy density in terms of three integrals:
\begin{widetext}
\begin{eqnarray}
\frac{\partial\beta\Delta f_A}{\partial\ell_B}&=&\frac{2}{\ell_B}\,\frac{\partial\kappa}{\partial\ell_B}\,\int\,\frac{d^2q}{(2\pi)^2}\,\left\{{\mathcal{I}}_1+2{\mathcal{I}}_2\right\}-\frac{4}{\ell_B}\,\frac{\kappa^2}{\sqrt{1+(\kappa\lambda)^2}}
\frac{\partial\lambda}{\partial\ell_B}\,\int\,\frac{d^2q}{(2\pi)^2} \,{\mathcal{I}}_3\;,
\label{deltaF2}
\end{eqnarray}
where
\begin{eqnarray}
\label{I1}
\mathcal{I}_1&\equiv&\kappa\int^{\infty}_0 dz\,\left[G(z,z;q)-G_{DH}(z,z;q)\right]\;,\\
\label{I2}
\mathcal{I}_2&\equiv&\kappa\int^{\infty}_0 dz\,\left[G(z,z;q)-G_{DH}(z,z;q)\right]{\rm{csch}}^2\kappa\left(z+z_0\right)\left[1-\kappa(z+s)\,{\rm{coth}}\kappa(z+z_0)\right]\;,\\
\mathcal{I}_3&\equiv&\kappa\int^{\infty}_0 dz\,\left[G(z,z;q)-G_0(z,z;q)\right]{\rm{csch}}^2\kappa(z+z_0){\rm{coth}}\kappa(z+z_0)
\label{I3}
\end{eqnarray}
\end{widetext}
It remains now to evaluate these integrals.  We will present the details of which in the Appendix.  For now, we simply state that after massaging these integrals further and integrating over $\ell_B$, we would finally arrive at the one-loop correction,  induced by salt density fluctuation, to the free-energy density in terms of the dimensionless parameter $x=\kappa\lambda$:
\begin{widetext}
\begin{eqnarray}
\beta\Delta f_A&=&-\frac{\mathcal{K}(x)}{4\pi\lambda^2}
\label{final1}
\end{eqnarray}
where
\begin{eqnarray}
\mathcal{K}(x)&=&-\frac{\pi}{2\sqrt{3}}+\sqrt{3+2x^2-x^4}\tan^{-1}\frac{\sqrt{3-x^2}}{2x+\sqrt{1+x^2}}+2x\left(\sqrt{1+x^2}-x\right)+x^2\log\frac{\sqrt{1+x^2}}{x}\nonumber\\
&&-\frac{1-x^2}{2}\log\left(1+\frac{x}{\sqrt{1+x^2}}\right)\;.
\label{final2}
\end{eqnarray}
\end{widetext}

We note that the fluctuation-induced correction to the free-energy density has the overall effect of lowering the total free-energy density.  This is in line with our expectation for performing perturbation about a stable ground state.  It also makes sense that the one-loop correction would lower the free-energy density since mean-field theory usually overestimates repulsive interactions due to the neglect of correlations.  A plot of $\mathcal{K}(x)$ is shown in Fig. (\ref{K}).  We see that the one-loop correction to the free energy density saturates quickly when $x>2$ in the high-salt limit.  Therefore, in agreement with previous finding, in the regime where the interaction between salt ions at their mean separation is larger than thermal energy, we can simply define a renormalized screening length in the mean-field Poisson-Boltzmann description that results in a renormalized free-energy density.  And the difference of which from the bare mean-field free energy does not  depend sensitively at all on the exact value of the screening length in the high-salt regime.  However, fluctuation effects become important in systems with low to intermediate salt-density characterized by $0<x<2$, and whose correction to the free energy density depends sensitively on the salt density as well as their charge valency.

\begin{figure}[htb]
\vspace{2mm}
\hspace{23mm}
\includegraphics{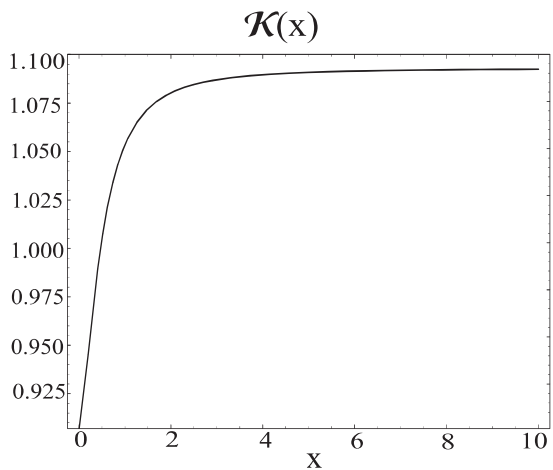}
\vspace{2mm}
\caption{Plot of $\mathcal{K}(x)$ as a function of $x=\kappa\lambda$.}
\label{K}
\end{figure}
\section{Conclusion}

We have formulated in a field-theoretic language the problem of a charged plate in a generally asymmetric salt solution containing both counterions and coions.  We have obtained the non-linear Poisson-Boltzmann equation as a saddle point of our field-theory action.  Specializing to the case of symmetric salt ions and using the full nonlinear mean-field solution,  we systematically incorporate the effect of fluctuation in ion densities and computed the one-loop correction to the free-energy density in closed form by integrating a Gaussian in the action and using the techniques of Green's function.  We have found that for systems with low to moderate salt density, fluctuation corrections to the free-energy depends sensitively on the salt concentration as well as their charge valency.  We have also considered the interaction of two point charges near the charged plate and found that density fluctuation suppresses repulsion between two point-charges.  Therefore fluctuation in charge densities reduces the screening effect of the salt ions as well.  

Consequently, we see that the mean-field Poisson-Boltzmann approach may have wider applicability than we originally anticipate.  Even in systems with high salt density or high-valency ions, the incorporation of their density fluctuation actually offsets the supposed stronger correlation between these ions, for which one would naively expect to invalidate the mean-field approach.  Thus in principle,  using a suitably chosen "renormalized" screening length, mean-field predictions can be made to fit properties of these charged systems in the high salt limit.  We note that in contrast to the perturbation carried out in the strong-coupling regime~\cite{Netz00b},  the one-loop corrections that we have obtained are proportional the the surface charge density.  Therefore, a high charge density for the plate should create a similar effect seen the high-salt limit.

So far we have only considered the effects of electrolyte fluctuation on the electrostatic interaction.   In reality, other factors such as the van der Waals forces are also essential to the physics of these systems.  Though often practiced in calculations, the separation of electrostatic interactions from van der Waals ones is rather artificial, since quantum-mechanically it is the fluctuation in the local electrostatic potential that gives rise to these attractive dispersion forces.  It has been shown~\cite{Lifshitz, Ninham, Mitchell} that the zero-frequency van der Waals interactions are altered significantly by the presence of salt ions, while the nonzero-frequency dispersion can be effectively taken as independent of ionic densities since they do not respond to high-frequency fields.  In regimes where linear superposition is applicable, our methodology and results for the one-plate case can be useful in studying the static contribution to these attractive dispersion forces.  Our expression for the one-loop free energy should also provide a more accurate description of the electrostatics in the study of phase transitions of a variety of systems that involve charged components in aqueous solutions. 

In conclusion, it is our hope that our solution would provide some insights as to why the mean-field Poisson Boltzmann approach seems to work in situations where the assumption of mean field should not be applicable~\cite{Travesset05, Schiessel2003}.  It is our further hope that by means of such techniques as the Derjaguin approximation and superposition principle, our solution can be applied to provide better approximation even to the more general cases involving spherical symmetry or multiple charged plates in salt solution.  The system of two charged plates in salt solution is of particular interest given the possible existence of attractive electrostatic forces between them.  Our results for this case will be published elsewhere~\cite{Kung2006}.
\\
\\

\acknowledgments
W.~K.  acknowledges support from NSF Grant DMR03-05407 and DMR02-19292.

\appendix

\section{Evaluation of integrals ${\mathcal{I}}_1, {\mathcal{I}}_2, {\mathcal{I}}_3$}

In the appendix, we will present a few details on the evaluation of the integrals ${\mathcal{I}}_1, {\mathcal{I}}_2, {\mathcal{I}}_3$ found in section IV-B, as follows:
\begin{widetext}
\begin{eqnarray}
\frac{\partial\beta\Delta f_A}{\partial\ell_B}&=&\frac{2}{\ell_B}\,\frac{\partial\kappa}{\partial\ell_B}\,\int\,\frac{d^2q}{(2\pi)^2}\,\left\{{\mathcal{I}}_1+2{\mathcal{I}}_2\right\}-\frac{4}{\ell_B}\,\frac{\kappa^2}{\sqrt{1+(\kappa\lambda)^2}}
\frac{\partial\lambda}{\partial\ell_B}\,\int\,\frac{d^2q}{(2\pi)^2} \,{\mathcal{I}}_3\;,
\label{deltaF2}
\end{eqnarray}
where
\begin{eqnarray}
\label{I1}
\mathcal{I}_1&\equiv&\kappa\int^{\infty}_0 dz\,\left[G(z,z;q)-G_{DH}(z,z;q)\right]\;,\\
\label{I2}
\mathcal{I}_2&\equiv&\kappa\int^{\infty}_0 dz\,\left[G(z,z;q)-G_{DH}(z,z;q)\right]{\rm{csch}}^2\kappa\left(z+z_0\right)\left[1-\kappa(z+s)\,{\rm{coth}}\kappa(z+z_0)\right]\;,\\
\mathcal{I}_3&\equiv&\kappa\int^{\infty}_0 dz\,\left[G(z,z;q)-G_0(z,z;q)\right]{\rm{csch}}^2\kappa(z+z_0){\rm{coth}}\kappa(z+z_0)
\label{I3}
\end{eqnarray}
\end{widetext}
To integrate $z$ over the range $[0,\infty)$, we first substitute Eqs. (\ref{selfenergy}) and (\ref{DH}) into Eqs. (\ref{I1})-(\ref{I3}), and we repeatedly apply integration by parts to obtain the following results, in terms of $x=\kappa\lambda$ and $\alpha=\frac 12\sqrt{1+\frac{q^2}{\kappa^2}}$:
\begin{widetext}
\begin{eqnarray}
{\mathcal{I}}_1&=&-A(q)\left(\frac{\sqrt{1+x^2}}{x}-1\right)+\frac{B(q)}{4\alpha}\left(1+\frac{1}{\alpha}\frac{\sqrt{1+x^2}}{x}+\frac{1}{4\alpha^2}\right)\;,\\
{\mathcal{I}}_2&=&\frac{\ell_B}{4}\left(\frac{1}{\sqrt{\kappa^2+q^2}}-\frac{1}{q}\right)\left(\frac{x}{\sqrt{1+x^2}}-1\right)-A(q)\left[-\frac 12\left(\frac{\sqrt{1+x^2}}{x}-1\right)+\frac 14\frac{1}{x\sqrt{1+x^2}}\right]\;,\\
{\mathcal{I}}_3&=&\frac{\ell_B}{4x^2}\left(\frac{1}{\sqrt{\kappa^2+q^2}}-\frac{1}{q}\right)-\frac{A(q)}{4x^4}+B(q)\left[\frac{1}{4\alpha}\frac{\sqrt{1+x^2}}{x^3}+\frac{1}{16\alpha^2\,x^2}\left(2+\frac{1}{x^2}\right)\right]\;,
\end{eqnarray}
\end{widetext}
where
\begin{eqnarray}
A(q)&=&\frac{\ell_B}{2q^2}\frac{\kappa^2}{\sqrt{\kappa^2+q^2}}\;,\\
B(q)&=&\frac{\ell_B}{2q^2}{\sqrt{\kappa^2+q^2}}\,M(q)\;.
\end{eqnarray}
To compute the momentum integrations, we use the standard method of converting to polar coordinates in Fourier space.  Repeated application of partial-fraction decomposition in integrals involving $M(q)$ and doing some algebraic manipulations lead to the following expressions:
\begin{widetext}
\begin{eqnarray}
\label{A6}
\int\frac{d^2q}{(2\pi)^2}\left({\mathcal{I}}_1+2{\mathcal{I}}_2\right)&=&-\frac{\ell_B\,\kappa}{4\pi}\left(\frac{x}{\sqrt{1+x^2}}-1\right)-\frac{\ell_B}{8\pi}\frac{\kappa}{1+x^2}\int^{\infty}_{\frac{x}{\sqrt{1+x^2}}}\frac{dy}{y}\frac{1+y}{\frac{1}{1+x^2}+y+y^2}\;,\\
\int\frac{d^2q}{(2\pi)^2}\,{\mathcal{I}}_3&=&-\frac{\ell_B\,\kappa}{8\pi\,x^2}-\frac{\ell_B\,\kappa}{16\pi\,x^2\lambda}\frac{1}{\sqrt{1+x^2}}\int^{\infty}_{\frac{x}{\sqrt{1+x^2}}}\frac{dy}{1+y}\frac{2+y}{\frac{1}{1+x^2}+y+y^2}\;,
\label{A7}
\end{eqnarray}
\end{widetext}
where $y=\frac{x\lambda}{\sqrt{1+x^2}}$.  Using Eqs. (\ref{A6}) and (\ref{A7}), we can now integrate away $\ell_B$ on both sides of Eq. (\ref{deltaF2}) and perform the $\kappa$- and $\lambda$-integrations, respectively, in the first and second term of Eq. (\ref{deltaF2}).  Upon further simplifying, we obtain the following intermediate expression for the one-loop free-energy density:
\begin{widetext}
\begin{eqnarray}
\beta\Delta f_A&=&-\frac{1}{4\pi\lambda^2}\left[x\left(\sqrt{1+x^2}-x\right)-\sinh^{-1}x\right]-\frac{1}{4\pi\lambda^2} {\mathcal{I}}_A-\frac{1}{2\pi\lambda^2}x\sqrt{1+x^2}+\frac{\kappa^2}{4\pi}{\mathcal{I}}_B\;,
\label{A8}
\end{eqnarray}
where
\begin{eqnarray}
\label{A9}
 {\mathcal{I}}_A&=&\int \frac{x\,dx}{1+x^2}\int^{\infty}_{\frac{x}{\sqrt{1+x^2}}}\frac{dy}{y}\frac{1+y}{\frac{1}{1+x^2}+y+y^2}\;,\\
  {\mathcal{I}}_B&=&\int \frac{dx}{x^3}\frac{1}{1+x^2}\int^{\infty}_{\frac{x}{\sqrt{1+x^2}}}\frac{dy}{1+y}\frac{2+y}{\frac{1}{1+x^2}+y+y^2}\;,
\label{A10}
\end{eqnarray}
\end{widetext}
Evaluation of the remaining two integrals in Eqs. (\ref{A9}) and (\ref{A10}) are straightforward.  Substituting the subsequent results back in Eq. (\ref{A8}) and simplifying lead us to the final expression of the one-loop correction to the free-energy density in Eqs. (\ref{final1}) and (\ref{final2}).

\end{document}